\documentclass[10pt, conference]{IEEEtran}
\usepackage{geometry}
\usepackage{bbold}
\usepackage{float}
\usepackage{stfloats}
\usepackage{subcaption}%
\usepackage[cmex10]{amsmath}
\usepackage{amssymb}
\usepackage{verbatim}
\usepackage{array}
\usepackage{latexsym}
\usepackage{color}
\usepackage{tabularx} 
\usepackage{textgreek}
\usepackage{subscript}
\usepackage{rotating}
\usepackage{booktabs,multirow}
\usepackage{mdframed}	
\usepackage{tabularx}
\usepackage{amsmath}
\usepackage{makecell}
\usepackage{tabto}
\usepackage{pifont}
\newcommand{\cmark}{\textcolor{green}{\ding{51}}} 
\newcommand{\xmark}{\textcolor{red}{\ding{55}}}   
\usepackage{notoccite}
\usepackage{mathrsfs}
\usepackage{url}
\usepackage{caption}
\usepackage{cite}
\usepackage{listings}
\usepackage{array}
\usepackage[table]{xcolor}

\definecolor{pinkpurple}{rgb}{0.1, 0.3, 0.9} 

\newcommand{\nref}[1]{\textcolor{red}{[{\bf NEED CITATION}]}}

\begin{document}

\setlength{\columnsep}{0.2 in}
\def\BibTeX{{\rm B\kern-.05em{\sc i\kern-.025em b}\kern-.08em T\kern-.1667em\lower.7ex\hbox{E}\kern-.125emX}}

\title{Meeting Deadlines in Motion: Deep RL for Real-Time Task Offloading in Vehicular Edge Networks}

\author{
	Mahsa~Paknejad,~Parisa~Fard~Moshiri,~Murat~Simsek~\IEEEmembership{Senior~Member,~IEEE},\\~Burak Kantarci,~\IEEEmembership{Senior Member,~IEEE}~and~Hussein~T.~Mouftah,~\IEEEmembership{Fellow,~IEEE}
\thanks{
The authors are with the School of Electrical Engineering and Computer Science at the University of Ottawa, Ottawa, ON, K1N 6N5, Canada.
        E-mail: \{mahsa.paknejad, parisa.fard.moshiri, murat.simsek, burak.kantarci, mouftah\}@uottawa.ca}  
}

\maketitle
\thispagestyle{empty}
\pagestyle{empty}
\begin{abstract}
Vehicular Mobile Edge Computing (VEC) drives the future by enabling low-latency, high-efficiency data processing at the very edge of vehicular networks. This drives innovation in key areas such as autonomous driving, intelligent transportation systems, and real-time analytics. Despite its potential, VEC faces significant challenges, particularly in adhering to strict task offloading deadlines, as vehicles remain within the coverage area of Roadside Units (RSUs) for only brief periods. To tackle this challenge, this paper evaluates the performance boundaries of task processing by initially establishing a theoretical limit using Particle Swarm Optimization (PSO) in a static environment. To address more dynamic and practical scenarios, PSO, Deep Q-Network (DQN), and Proximal Policy Optimization (PPO) models are implemented in an online setting. The objective is to minimize dropped tasks and reduce end-to-end (E2E) latency, covering both communication and computation delays.
Experimental results demonstrate that the DQN model considerably surpasses the dynamic PSO approach, achieving a 99.2\% reduction in execution time. Furthermore, It leads to a reduction in dropped tasks by 2.5\% relative to dynamic PSO and achieves 18.6\% lower E2E latency, highlighting the effectiveness of Deep Reinforcement Learning (DRL) in enabling scalable and efficient task management for VEC systems.

 
\end{abstract}

\IEEEpeerreviewmaketitle

\section{Introduction}

Mobile Edge Computing (MEC) significantly enhances vehicular task offloading by enabling vehicles to delegate complex computations to nearby edge servers, rather than relying solely on onboard systems or distant cloud infrastructure \cite{start}.
By supporting real-time data analysis, MEC helps vehicles operate more intelligently within their surroundings \cite{nof1}. This integration of edge computing with connected vehicles forms the foundation for Ubiquitous Intelligence, where smart decision-making happens seamlessly across the network, contributing to safer and more adaptive transportation ecosystems \cite{nof2}. Despite the advantages MEC brings to vehicular task offloading, several challenges remain, particularly in meeting the strict latency and deadline constraints of vehicle-based applications. One major issue is ensuring that offloaded tasks are not only processed quickly but also returned to the vehicle before it moves out of the coverage range of a Roadside Unit (RSU) \cite{mahsaicc}. Since vehicles are highly mobile and only stay connected to an RSU for a short period, any delay in processing or communication can result in missed deadlines and dropped tasks. Reducing task drop rates is therefore critical. 
One approach is to use optimization techniques such as Particle Swarm Optimization (PSO) \cite{mahsaicc}. These methods can help determine the most efficient order in which tasks should be assigned and processed by edge servers, aiming to minimize task drop rates \cite{parisa1}, \cite{parisaicc}. However, a key limitation is that such optimization algorithms often require significant computational time, making them less suitable for real-time scenarios where rapid decisions are critical. In particular, the dynamic version of PSO tends to cause a higher task drop rate, primarily due to its high execution time. In contrast, Machine Learning (ML) models offer a promising alternative for real-time task offloading. Once trained, ML models can quickly predict optimal or near-optimal task assignments. Building on the advantages of ML, Reinforcement Learning (RL) can be even more effective \cite{intro3}. Unlike traditional ML models, RL focuses on learning optimal policies for sequential decision-making, rather than just predicting outcomes. It can improve upon suboptimal behaviors in the data by reasoning about long-term effects of actions, making it ideal for complex tasks where decisions impact future outcomes.
This paper makes the following key contributions:
\begin{enumerate}
\item We establish a theoretical limit for task offloading performance in vehicular networks by implementing PSO in a static and ideal environment.
\item We present a dynamic environment that utilizes a decision window mechanism to manage incoming tasks prior to MEC availability. The proposed approach prioritizes reducing the number of dropped tasks while minimizing the end-to-end (E2E) latency. Decision-making within the window relies on a reward function in RL models and on dynamic PSO.
\item A novel dynamic reward function is employed during the training of Deep Q-Network (DQN) and Proximal Policy Optimization (PPO). The resulting best policy is then tested on a new dataset, eliminating the need for retraining across different datasets.

\end{enumerate}
In the following sections, Section II reviews related work, Section III presents the proposed methodology, Section IV discusses the performance evaluation, and Section V provides the conclusion.

\begin{table*}[!t]
\caption{Assessment of our work in relation to existing research}

\centering
\renewcommand{\arraystretch}{1.7} 
\newcolumntype{M}[1]{>{\centering\arraybackslash}m{#1}}
\begin{tabular}{|M{0.7cm}|M{1.5cm}|M{1.1cm}|M{1.3cm}|M{0.9cm}|M{0.9cm}|M{0.7cm}|c|M{0.7cm}|M{5cm}|}
\hline

\textbf{Paper} & \textbf{Theoritical Limit} & \textbf{Dynamic} & \textbf{MTG} & \multicolumn{2}{c|}{\textbf{Training}} & \multicolumn{3}{c|}{\textbf{KPI}} & \textbf{Algorithm} \\ \cline{5-6} \cline{7-9}
               &         &          &                         &  \textbf{Offline} & \textbf{Online} &  \textbf{AET} & \textbf{Latency} & \textbf{DTN} & \\
\hline

\cite{relate1}  & {\xmark} & {\cmark} & - & - & - & {\xmark} & {\cmark} & {\xmark}  & PSO, ICRI  \\ \hline

\cite{relate2}  & {\xmark} & {\cmark} & MATLAB & - & - & {\xmark} & {\cmark} & {\xmark}  & OJTR, HJTR  \\ \hline

\cite{relate3}  & {\xmark} & {\cmark} & Python & - & {\cmark} & {\xmark} & {\cmark} & {\xmark}  & Deep Meta-RL, DQN \\ \hline


\cite{relate5}  & {\xmark} & {\cmark} & - & {\cmark} & - & {\xmark} & {\cmark} & {\xmark}  & AODAI, ACTO-n, DQN, SARSA   \\ \hline

\cite{relate7}  & {\xmark} & {\cmark} & Python & - & {\cmark} & {\xmark} & {\cmark} & {\xmark}  & DDQN, ALTO, UCB, AdaUCB \\ \hline

\cite{relate6}  & {\xmark} & {\cmark} & DAGGEN & {\cmark} & - & {\cmark} & {\cmark} & {\xmark}  & SMRL-MTO, Policy Gradient, PPO, DQN \\ \hline

\cite{relate8}  & {\xmark} & {\cmark} & SUMO, OMNet++ & - & {\cmark} & {\xmark} & {\cmark} & {\xmark}  & COADRL, HGOS, TOSO, TOSM, TNP \\ \hline

Ours  & {\cmark} & {\cmark} & SUMO & {\cmark} & - & {\cmark} & {\cmark} & {\cmark}  &  Dynamic PSO, DQN, PPO \\ \hline

\end{tabular}

\vspace{2mm}
\textit{* MTG: Mobility Trace Generator, KPI: Key Performance Indicator, AET: Algorithm Execution Time, DTR: Drop Task Ratio, DAGGEN : Directed Acyclic Graph Generator}
\label{tab:gaptable}
\end{table*}

\section{Related Work}

Extensive research has been conducted on task offloading in MEC to address its inherent limitations and enhance efficiency. However, a significant challenge persists in meeting the rapidly increasing demand for ultra-low-latency task processing in Internet of Vehicles (IoV) environments. To tackle this, optimization techniques such as PSO have been
explored. In one approach, PSO is used to jointly determine assistant vehicle selection and transmit power in a hybrid MEC for Vehicle-to-Everything (MEC-V2X) setting, where tasks can be offloaded not only to RSUs but also to nearby vehicles \cite{relate1}. 

In addition to optimization methods with high execution times, it's crucial to account for faster approaches. A recent study tackles this challenge by combining optimization techniques such as generalized benders decomposition and a low-complexity heuristic algorithm to efficiently minimize task processing delays \cite{relate2}. RL has also emerged as a powerful tool for task offloading in MEC due to its ability to make dynamic decisions in complex environments. Building on this, an online Deep Meta Reinforcement Learning (Meta-RL) framework models the offloading process as a Markov Decision Process to enhance adaptability in B5G/6G-enabled IoV systems and introduces a two-loop learning approach for fast decision-making \cite{relate3}. A recent approach introduces two novel RL-based methods, a value-based Advantage-Oriented Dueling Actor-Insulator Network (AODAI) and a policy-based Actor–Critic Task Offloading (ACTO-n) scheme\cite{relate5}.

A handover-enabled dynamic computation offloading framework further advances this direction by incorporating DRL with Double Q-learning to address the challenges of high mobility in vehicular edge computing environments. By enabling cooperation between smart vehicles and RSUs, the system dynamically selects suitable offloading targets based on environment feedback, thereby optimizing the latency. The proposed solution leverages experience replay and dual networks within a double deep Q-network (DDQN) architecture to make robust decisions in dynamic B5G/6G vehicular networks \cite{relate7}.
Furthermore, a Seq2Seq-based Meta-RL framework can be developed to tackle multi-task offloading (MTO) in dynamic environments. By framing the offloading process as a series of Markov Decision Processes (MDPs) and integrating a Bi-GRU encoder with attention mechanisms and a model-agnostic meta-learning (MAML) approach, this method enables rapid adaptation to changing task structures and MEC configurations \cite{relate6}.
In a similar direction, a Computation Offloading and Allocation with Deep Reinforcement Learning (COADRL) framework is proposed to address dynamic computation offloading in heterogeneous vehicular networks. By modeling stochastic task arrivals and time-varying channel conditions, this approach formulates offloading, migration, and bandwidth allocation decisions as part of a DRL-based optimization process. Using a DDQN architecture with experience replay, COADRL aims to minimize overall costs while adaptively maximizing long-term rewards without prior knowledge of the environment \cite{relate8}.
While these studies provide valuable insights, many do not provide comparisons to a theoretical limit and the reporting of execution times, both of which are critical for real‑time decision‑making. Another important consideration often overlooked is the number of dropped tasks and the objective of minimizing them. \tablename~\ref{tab:gaptable} compares our approach with existing studies, highlighting the gaps that this paper addresses.

Our work builds upon our previous research \cite{mahsaicc}, extending it to dynamic environments. We focus on intelligently assigning tasks based on waiting times, with the goal of minimizing overall latency and task drops, a metric that is often overlooked in existing studies. We deploy PSO, DQN, and PPO within a dynamic framework, demonstrating that RL models not only achieve significantly lower execution times compared to PSO, but also deliver performance close to the theoretical limit.

\section{Methodologies under Study}

\begin{figure*}[!hbt]
        \centering
        \includegraphics[width = 0.9\textwidth, trim=0cm 0cm 0cm 0cm,clip]{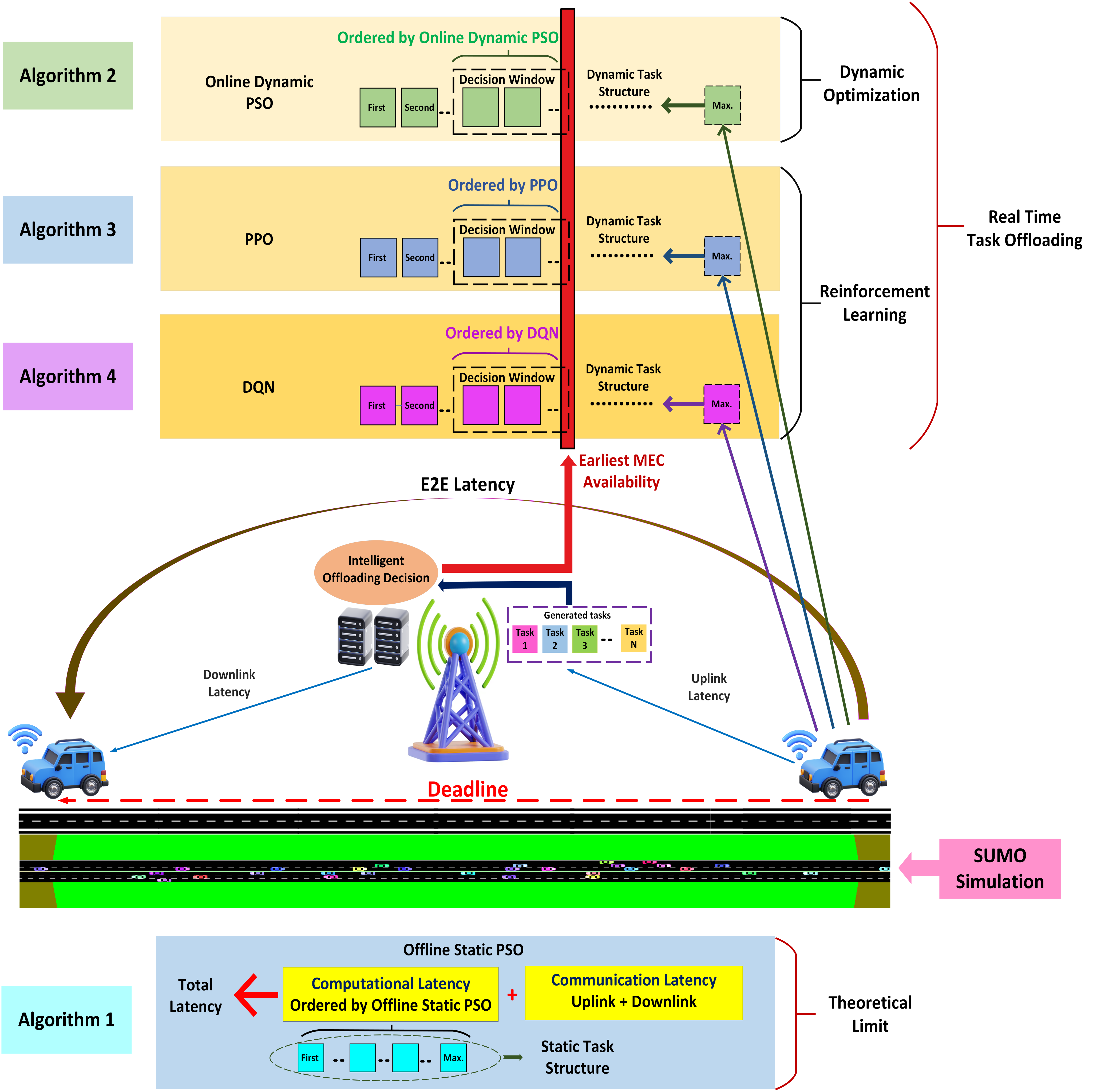}
        \caption{Overview of Vehicular Edge Computing for Static and Dynamic Task Offloading}
        \label{fig:scenario} 
\end{figure*}

\subsection{The Offloading Schemes}

Based on \figurename~\ref{fig:scenario}, we begin with an offline and static variant of PSO, referred to as Off-Sta-PSO, which runs on the entire mobility traces generated by the Simulation of Urban Mobility (SUMO). This model represents a theoretical limit, as it operates under the assumption of a static and ideal environment. In this static scenario, PSO's execution time is not considered, and tasks are handled offline, meaning there is prior knowledge of all incoming tasks at the RSU.

To explore more practical and dynamic scenarios, we introduce On-Dyn-PSO, which applies PSO in an online and dynamic environment. This means that tasks are handled in real time as they arrive, without any prior knowledge. This dynamic decision-making is important, as vehicles spend only a short time within the RSU's range. However, due to PSO's long execution time, it is inefficient for real-time applications. To address this limitation, we implement both off-policy and on-policy RL algorithms, using DQN and PPO, respectively.

In the dynamic setting, tasks that arrive before the earliest availability time (the moment when at least one MEC becomes available) are collected into a queue , named  decision window. For each window, the dynamic model determines which task should be assigned to the available MEC, aiming to minimize both the number of dropped tasks and the E2E latency.

After each assignment, the earliest availability time is updated based on the status of all MECs. If a task arrives when a MEC is already available, it is assigned immediately. In such simple cases, intelligent decision-making is unnecessary, which simplifies the offloading process and reduces execution time. The DQN and PPO models are first trained (referred to as the Baseline model) on randomly generated SUMO mobility traces. Once training is complete, the best policy is saved and later applied to the actual target dataset without further retraining; this version is called the Test-only model. This approach helps evaluate an RL model's ability to generalize and perform under real-world conditions, ensuring robustness and preventing overfitting. It also reduces training time and computational cost when deploying the model to new datasets.

\subsection{The mathematical framework for task offloading}

\begin{table}[!hbt]
\centering
\renewcommand{\arraystretch}{1.7} 
 
\caption{Notation table}
\label{tab:notations}
\begin{tabularx}
{0.43\textwidth}{|c|X|}
\hline
\textbf{Parameter} & \textbf{Description} \\
\hline
\(T_{i}^{\text{sp}}\) & Start processing time for task \( i \) \\
\hline
\(T_{i}^{\text{p}}\) & Processing time for task \( i \)\\
\hline
\(T_{i}^{\text{w}}\) & Waiting time for task \( i \)\\
\hline
\(T_{i}^{\text{ar}}\) & Arrival time of task \( i \) on the RSU\\
\hline
\(L_{i}^{\text{p}}\) & Computation Latency of task \( i \)\\
\hline
\( S_i \) & Size of task \( i \)\\
\hline
\( r_i \) & transmission rate for task \( i \)\\
\hline
\( T_i^{\text{cm}} \) & Communication time for task \(i\)\\
\hline
\(L_{i}^{\text{e2e}}\) & End-to-end latency for task \( i \)\\
\hline
\(x_{ij}\) & Binary decision variable for assigning task \(i\) to MEC \(j\)\\
\hline
\(D\) & Number of dropped tasks\\
\hline
\(N_{}^{\text{}}\) & Total number of tasks\\
\hline
\(M_{}^{\text{}}\) & Total number of MEC servers\\
\hline
\( T_{j}^{\text{av}} \) & The time when MEC \(j\) becomes available\\
\hline
\( T_{e}^{\text{av}} \) & Earliest MEC availability time \\
\hline

\(T_{yl}^{\text{ar}}\) & Arrival time of task \(l\) in subset \(y\)\\
\hline
\(T_{yl}^{\text{D}}\) & The deadline of task \(l\) in subset \(y\)\\
\hline
\(T_{yl}^{\text{R}}\) & Remaining time in the RSU range for task \(l\) in subset \(y\)\\
\hline
\(T_{yl}^{\text{p}}\) & Processing time required for task \(l\) in subset \(y\)\\
\hline
\({g_{y,l}^{\text{d}}}\) &
Time gap related to task \(l\) in subset \(y\)\\
\hline
\(R_{y,l}^{\mathrm{L}}\) &  Latency-based reward for assigning task \(l\) in subset \(y\)\\
\hline
\(R_{y,l}^{\mathrm{d}}\) &  Drop-task reward for assigning task \(l\) in subset \(y\)\\
\hline
\(R_{y,l}^{\mathrm{T}}\) &  Total reward for assigning task \(l\) in subset \(y\)\\
\hline
\(T\) & The complete collection of tasks in the simulation\\
\hline
\(S\) & The collection of tasks within a decision window\\
\hline
\(Q_y\) & The subset of tasks within a decision window that can be processed and returned before their deadlines\\
\hline
\(s_t\) & The state of the system at time \(t\)\\
\hline
\(a_y\) & The action taken at subset \(y\)\\
\hline

\(\mathcal{A}(y)\) & The complete set of possible actions for subset \(y\)\\

\hline
\(P(y)\) & The processing time for all tasks within subset \(y\)\\

\hline
\end{tabularx}
\label{tab:notations} 
\end{table} 

\subsubsection{Computation Latency}
Since we have \(M\) MEC servers, the first \(M\) tasks are assigned to a server and begin processing as soon as they arrive at the RSU, denoted by \(T_{i}^{\text{ar}}\), while the start processing time \( T_{i}^{\text{sp}} \) for the remaining tasks is set to the moment their preceding task has finished processing. Furthermore, the task processing times \(T_{i}^{\text{p}}\) correspond to the inference times reported in \cite{inference}. The notations employed in the mathematical formulas are detailed in \tablename~\ref{tab:notations}.


Most tasks experience a waiting time \(T_{i}^{\text{w}}\) at the RSU before being processed. Based on (\ref{eq:Lcp}), the computation latency \(L_{i}^{\text{p}}\) for each task is the sum of its processing time and waiting time.

\begin{equation}
L_{i}^{\text{p}} = T_{i}^{\text{p}} + \underbrace{(T_{i}^{\text{sp}} - T_{i}^{\text{ar}})}_{T_{i}^{\text{w}}}
    \label{eq:Lcp}
\end{equation}


\subsubsection{Communication Latency}



During transmission, if multiple tasks within the RSU’s range are offloaded simultaneously, the available bandwidth is shared among them. In such cases, the bandwidth allocated to each task is proportional to its size \cite{mahsaicc}. However, if a task is transmitted individually, it can utilize the entire available bandwidth. Based on our study in \cite{mahsaicc}, sharing the bandwidth can slightly reduce transmission latency compared to using a fixed bandwidth allocation.
The set of concurrent tasks is defined as:

\begin{equation}
     \mathcal{N}
  \;=\;
  \bigl\{\,i=1,\dots,N' \;\big|\;\mathcal{T}_{i} = \mathcal{T}\bigr\}
  \,
\end{equation} $\mathcal{T}_{i}$ denotes the offloading time when task \(i\) becomes ready for transmission, while $\mathcal{T}_{}$ represents the offloading time at which all tasks in \(\mathcal{N}\) are ready to be offloaded. Additionally, \(N'\) represents the total number of tasks in set \(\mathcal{N}\).
The bandwidth assigned to task \(i\) is calculated using (\ref{eq:B}). This bandwidth is further used to calculate the transmission rate, \(r_i\), which is needed to determine the transmission time.

\begin{equation}
B_{i}^{\text{}} = 
\begin{cases} 
B_{\text{max}} \cdot \frac{S_{i}^{\text{}}}{\sum_{i=1}^{N'} S_{i} }&  {N'>1}  \\ 
B_{\text{max}}  & {N'=1}
\end{cases}
    \label{eq:B}
\end{equation}

Where \(S_{i}\) represents the size of task \(i\).
The data rate, \(r_i\), is defined in (\ref{eq:r}), with \(p\) indicating the transmission power, \(g\) referring to the channel gain, and \(n_0\) denoting the noise power density.

\begin{equation}
r_i = B_i \cdot \log_2 \left( 1 + \frac{p \cdot g}{n_0} \right)
    \label{eq:r}
\end{equation}

To calculate the communication time \(T_i^{\text{cm}}\) we require this transmission rate \(r_i\). We assume that downlink communication is identical to uplink communication. As defined in (\ref{eq:end2end}), the E2E latency is the sum of the computation latency and both communication times \(T_i^{\text{cm}}\).






\begin{equation}
L_{i}^{\text{e2e}} = L_{i}^{\text{p}} + (2 \times \underbrace {\frac{S_i}{r_i}}_{T_{i}^{\text{cm}}})
    \label{eq:end2end}
\end{equation}

\subsubsection{Dropped Tasks}

A task is assigned to a MEC server only if its E2E latency does not exceed its deadline. If a task cannot meet this condition, it will be dropped. This assignment is represented by a binary variable \( x_{ij} \) as illustrated in (\ref{eq:assign}).


\begin{equation}
x_{ij} = 
\begin{cases} 
1 & \text{if task } i \text{ is assigned to MEC } j  \\ 
0 & \text{otherwise} 
\end{cases}
    \label{eq:assign}
\end{equation}


\subsubsection{Optimization Problem}
Our goal is to minimize dropped tasks \(D\) first, then reduce E2E latency for remaining tasks. The objective function (\ref{eq:obj}) balances these factors, with \(\lambda\) controlling the emphasis on latency and \((1-\lambda)\) prioritizing dropped task reduction. \(\lambda\) is set empirically through iterative testing to favor \(D\).

\vspace{-6mm}
{\small
\begin{equation}
\min \left( \lambda \left( \sum_{j=1}^{M} \sum_{i=1}^{N} L_{i}^{\text{e2e}} \cdot x_{ij} \right) + (1 - \lambda) \underbrace{\left( \frac{1}{N} \sum_{i=1}^{N} \left(1 - \sum_{j=1}^{M} x_{ij} \right) \right)}_{D} \right)
    \label{eq:obj}
\end{equation} 
}

According to constraint (\ref{eq:assign-con}), each task can be assigned to and processed by only one MEC server.

\begin{equation}
\sum_{j=1}^{M} x_{ij} \leq 1 \quad \forall i 
    \label{eq:assign-con}
\end{equation}

\subsubsection{Real-Time Approach}
To process tasks in real time, it is necessary to determine the availability of MEC servers after each assignment, which is computed according to (\ref{eq:ava}).

\begin{equation}
T_{j}^{\text{av}} = 
{(T_{i}^{\text{sp}} + T_{i}^{\text{p}}) \cdot x_{ij}}\quad \forall j 
\label{eq:ava}
\end{equation}





Let \( T = \{t_1, t_2, \dots, t_N\} \) represent the set of all tasks generated within the simulation. Once each task is assigned to a MEC server and all servers' availability times are updated, a decision window is formed. This decision window consists of all tasks that are waiting to be offloaded and is represented as
$S = \{t_{i_1}, t_{i_2}, \dots, t_{i_W}\}$ 
where W  is the total number of tasks in  S, while \(S \subseteq T\).
A subset of tasks \( Q_y \) within the decision window \( S \) is considered for assignment. These are tasks that arrive before the earliest availability and are not assigned yet. They are also those that can be processed and returned to the vehicles before their deadlines. If a task does not satisfy this condition, it is not included in the \(Q_y\) set. This subset is defined mathematically as:

\begin{align}
Q_y = \bigl\{t_{y_l} \in S \;\big|\; &\, 1 \leq l \leq W_y,\; T_{y_l}^{\text{ar}} \leq T_e^{\text{av}}, \notag \\
&\, T_{y_l}^w \leq T_{y_l}^R - T_{y_l}^p - T_{y_l}^{\text{cm}} \bigr\}
\end{align} where \(t_{y_l}\) is the task at position \(l\) in subset \(y\).

\vspace{5mm}
\subsubsection{Reinforcement Learning}
The definitions of state and action spaces in the RL model are as follows:

\paragraph{State Space}
It represents the current status of the system, which includes:

\begin{itemize}
    \item \( \{T_{j}^{\text{av}}\}_{j=1}^M \):  Availability times for all \( M \) MEC servers.
    \item \( \{T_{y_l}^{\text{ar}}, T_{y_l}^{\text{R}}, T_{y_l}^p\}_{y=1}^{W_s},\,_{l=1}^{W_y} \): Characteristics of tasks in the set of \( Q_y \).
\end{itemize}

At time \( t \), the system state is represented as:

\[
s_t = \Big\{\{T_{j}^{\text{av}}\}_{j=1}^M, \{T_{y_l}^{\text{ar}}, T_{y_l}^{\text{R}}, T_{y_l}^p\}_{y=1}^{W_s},\,_{l=1}^{W_y} \Big\}
\]

\paragraph{Action Space}

The action space defines the possible assignments for tasks in \(Q_y\). The action \(a_y\) selects which specific task from the subset \(Q_y\) should be assigned to the server. Mathematically, this is represented as:

\[
a_y \in \{1, 2, \dots, W_y\}, \quad
a_y = l \iff \text{assign task } t_{y_l}
\]


The full action space, denoted as \(\mathcal{A}(y)\), is given by:

\[
\mathcal{A}(y) = \{1, 2, \dots, W_y\}
\] where \(W_y\) represents the total number of tasks in the subset \(Q_y\).

\paragraph{Reward Function}

The reward mechanism in the RL model is composed of two complementary components, designed to balance two objectives, latency and dopped tasks.

\paragraph{Drop-Task Reward}

To encourage timely task assignment, a \emph{time gap} \(g_{y,l}^d\) is calculated for each task \(t_{y_l}\) in \(Q_y\). This is based on the difference between the task's deadline and the MEC server's earliest availability time:

\[
g_{y,l}^{\text{d}} = T_{y_l}^{\text{D}} - T_{e}^{\text{av}} \quad \forall y, l
\]

The total time gap for all tasks in \(Q_y\), denoted \(G_y^{\text{d}}\), is computed as:

\[
G_y^{\text{d}} = \sum_{l=1}^{W_y} g_{y,l}^{\text{d}} \quad \forall y
\]

Finally, the Drop-Task reward for assigning a specific task \(t_{y_l}\) at \(Q_y\) is defined as:

\[
R_{y,l}^{\text{d}} = \frac{100}{G_y^{\text{d}}} \cdot \bigl(G_y^{\text{d}} - \widetilde{g_{y,l}^{\text{d}}}\bigr) \quad \forall y, l
\] where \(\widetilde{g_{y,l}^{\text{d}}}\) represents the time gap associated with the task selected by the RL algorithm.


This formulation ensures that tasks with tighter deadlines relative to the MEC's availability are assigned a higher priority.

\paragraph{Latency-Based Reward} The cumulative processing time for all tasks in \(Q_y\), denoted as \(P_y\), is then:

\[
P_y = \sum_{l=1}^{W_y} T_{y_l}^p \quad \forall y
\]


The Latency-Based reward for assigning a specific task \(t_{y_l}\) at \(Q_y\) is formulated to encourage the minimization of latency:

\[
R_{y,l}^{\text{L}} = \frac{100}{P_y} \cdot \bigl(P_y - \widetilde{T_{y_l}^P}\bigr) \quad \forall y, l
\] where \(\widetilde{T_{y_l}^P}\) refers to the processing time of the task chosen by the RL algorithm. This formulation ensures that tasks with lower computational latencies are prioritized, promoting efficiency in the assignment process. 

\(R_{y,l}^{\mathrm{T}}\), the total reward for assigning \(t_{y_l}\) at \(Q_y\), is formulated as (\ref{eq:reward-total}).


\begin{equation}
  R_{y,l}^{\mathrm{T}}
  = R_{y,l}^{\mathrm{d}} + R_{y,l}^{\mathrm{L}}
  \quad \forall\,y,l
  \label{eq:reward-total}
\end{equation}

\begin{table}[!hbt]
\centering
\renewcommand{\arraystretch}{1.7} 
\caption{Parameters values} 
\begin{tabular}{|l|l|}
\hline
\textbf{Parameters} & \textbf{Value}   \\   \hline
Number of vehicles & 50, 100, 200   \\   \hline
Number of tasks per vehicle & 1 \\   \hline
Number of MEC servers & 2   \\   \hline
Max bandwidth & 20 Mhz   \\   \hline
Swarm size & 50 \\
\hline
$\lambda$ & 0.4 \\
\hline
Personal \& global learning coefficients & 1.49 \\
\hline
DQN learning rate & 0.0001 \\
\hline
PPO learning rates of actor and critic  & 0.0003 \\
\hline
DQN discount factor & 0.9 \\
\hline
PPO discount factor & 0.95 \\
\hline

\end{tabular}
\label{tab:parameters} 
\end{table}

\begin{figure}[!hbt]
    \centering
        \includegraphics[width=0.5\textwidth, trim=0cm 0.0cm 0cm 0cm, clip]{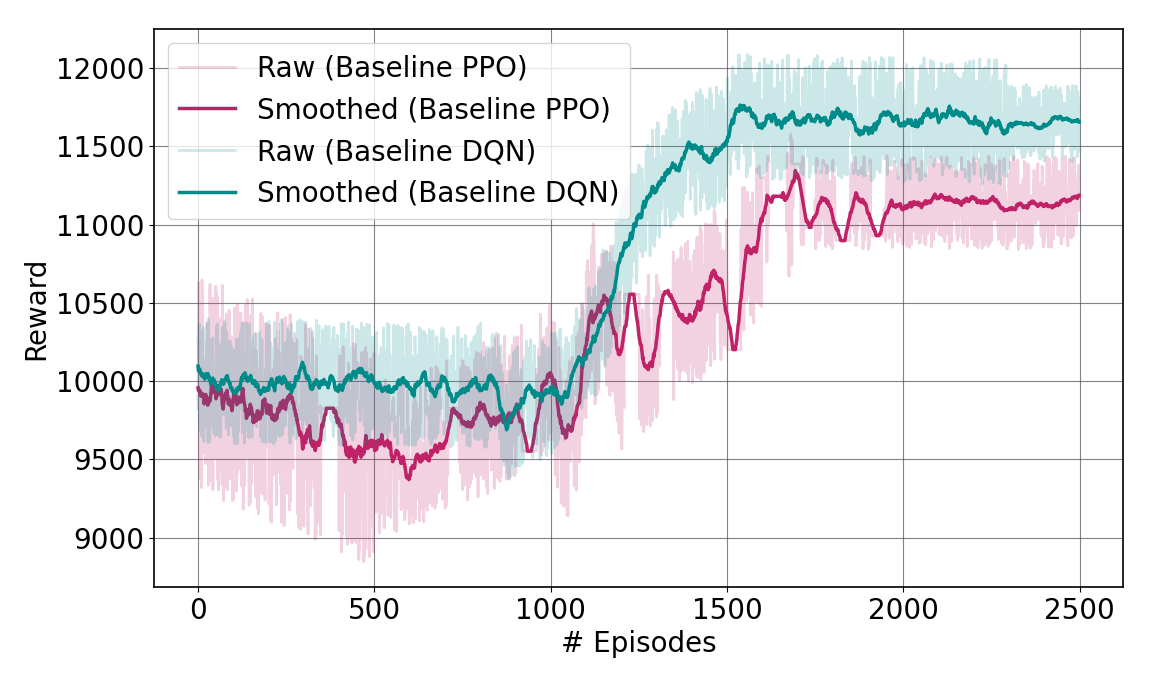}
        \caption{Reward Values for RL models}
        \label{fig:reward}
\end{figure}

\begin{figure}[!hbt]
    \centering
        \includegraphics[width=0.5\textwidth, trim=0.7cm 0.0cm 2cm 1.2cm, clip]{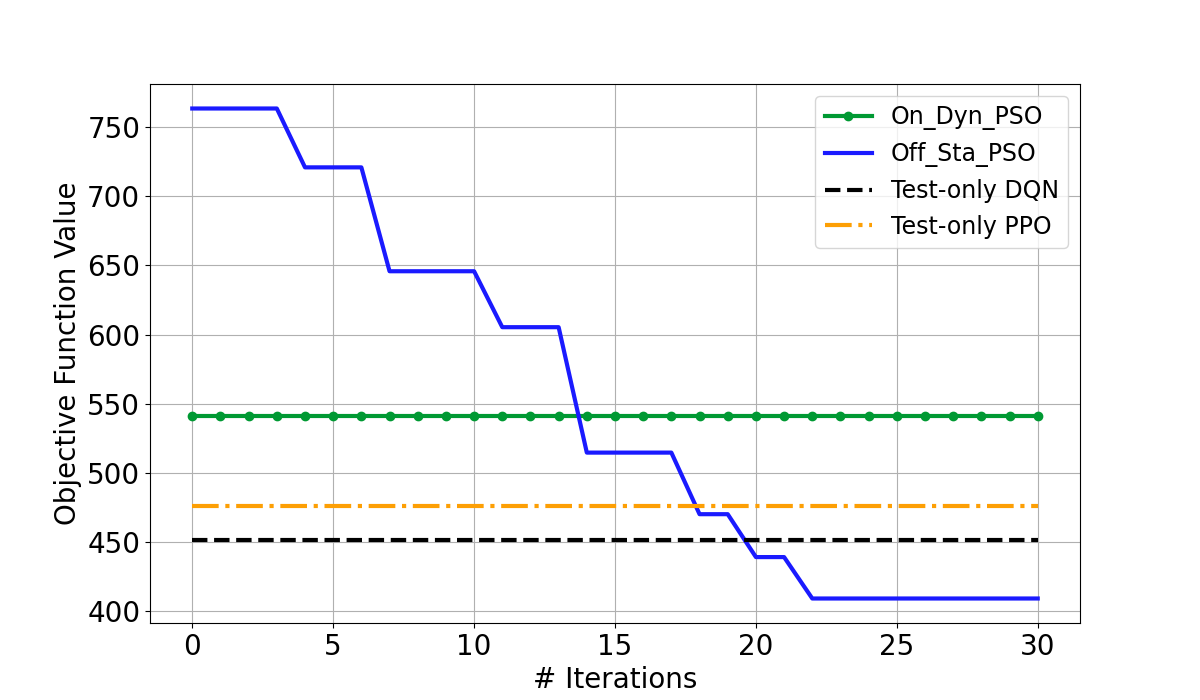}
        \caption{Objective Function Values for all algorithms}
        \label{fig:allobjective}
\end{figure}

\begin{table*}[h!]
 \caption{Total and Per-Window Execution Times for Test-only DQN and PPO}
    \centering
    \scalebox{0.87}{
    \renewcommand{\arraystretch}{1.9}
    \begin{tabular}{|c|c|c|c|c|c|c|c|c|c|c|}
        \hline
        \multirow{2}{*}{\textbf{Algorithm}} & \multicolumn{3}{|c|}{\textbf{Total Execution Time}} & \multicolumn{3}{|c|}{\textbf{Number of Decision Windows}} & \multicolumn{3}{|c|}{\textbf{Average Execution Time (s)}} \\ 
        \cline{2-10}
        & \textbf{50 Vehicles} & \textbf{100 Vehicles} & \textbf{200 Vehicles}
        & \textbf{50 Vehicles} & \textbf{100 Vehicles} & \textbf{200 Vehicles}
        & \textbf{50 Vehicles} & \textbf{100 Vehicles} & \textbf{200 Vehicles} \\
        \hline
        \textbf{Test-only DQN} & \textbf{0.28} & \textbf{1.6} & \textbf{10.62} & \textbf{20} & \textbf{62} & \textbf{178} & \textbf{0.014} & \textbf{0.026} & \textbf{0.06} \\
        \hline
        \textbf{Test-only PPO}  & \textbf{0.46} & \textbf{3.55} & \textbf{24.78} & \textbf{20} & \textbf{64} & \textbf{177} & \textbf{0.023} & \textbf{0.056} & \textbf{0.14} \\
        \hline
    \end{tabular}
    }
    \label{tab:exe}
\end{table*}

\begin{figure}[!hbt]
    \centering
\includegraphics[width=0.5\textwidth, trim=0.7cm 0.0cm 2cm 1.2cm, clip]{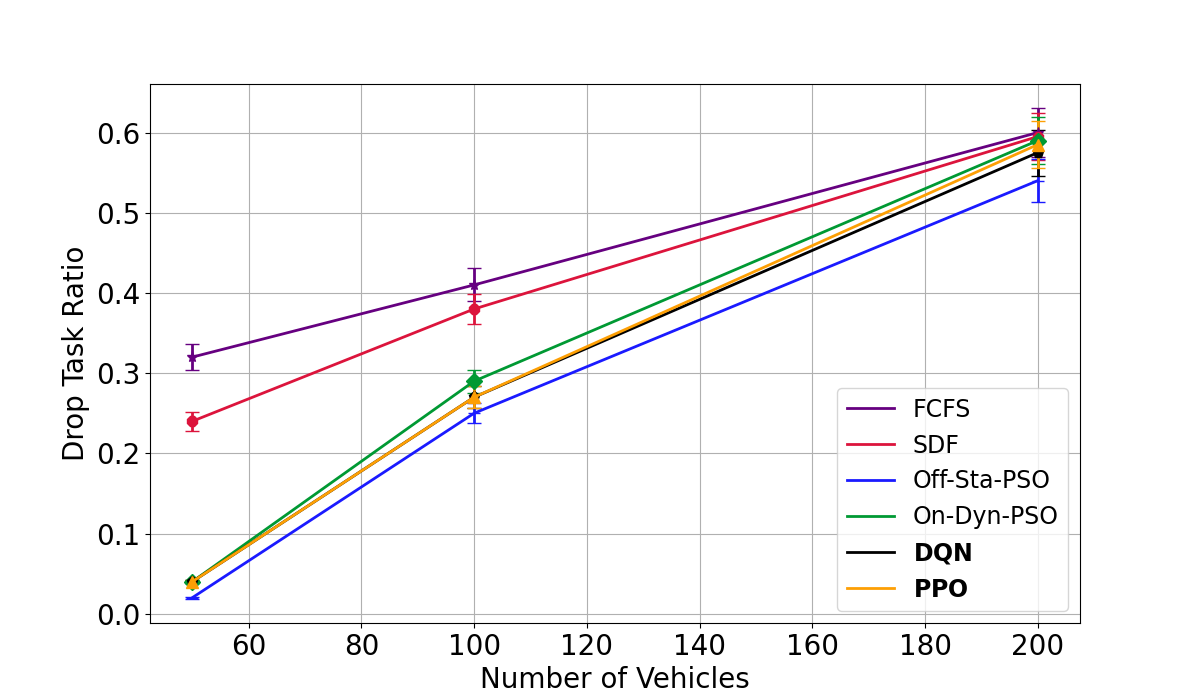}
    \caption{Average Drop Task Ratio}
    \label{fig:drop}
\end{figure}

\begin{figure}[!hbt]
    \centering
    \includegraphics[width=0.5\textwidth,trim=0.7cm 0.0cm 2cm 1.2cm, clip]{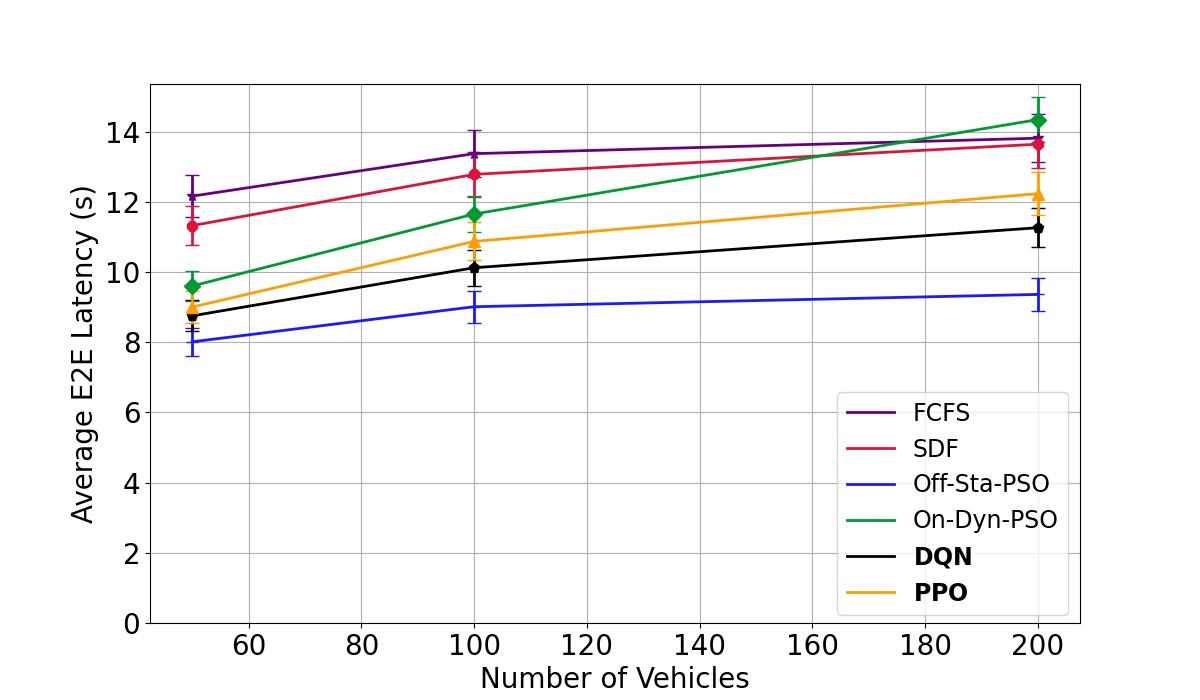}
    \caption{Average E2E Latency}
    \label{fig:latency}
\end{figure}

\begin{figure}[!hbt]
    \centering
    \includegraphics[width=0.5\textwidth, trim=0cm 0.0cm 2cm 1.2cm, clip]{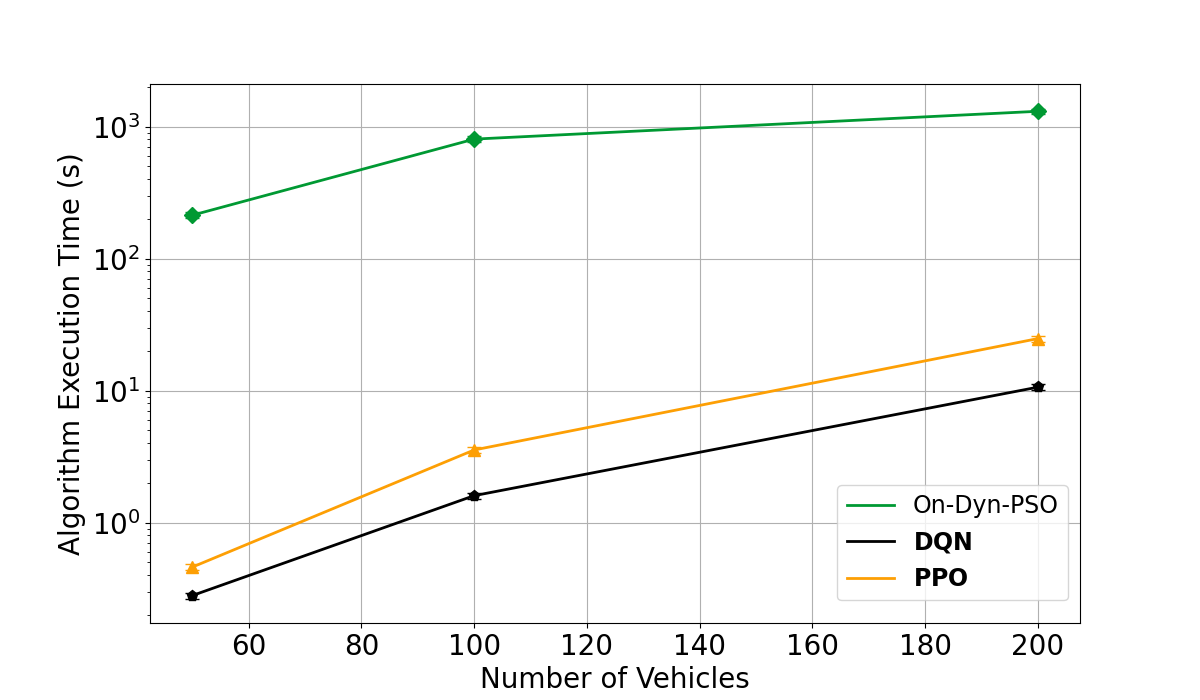}
    \caption{Logarithmic Algorithm Execution Time}
    \label{fig:execution}
\end{figure}

\begin{figure}[!hbt]
    \centering
    \includegraphics[width=0.5\textwidth, trim=1.2cm 0.0cm 2cm 1.2cm, clip]{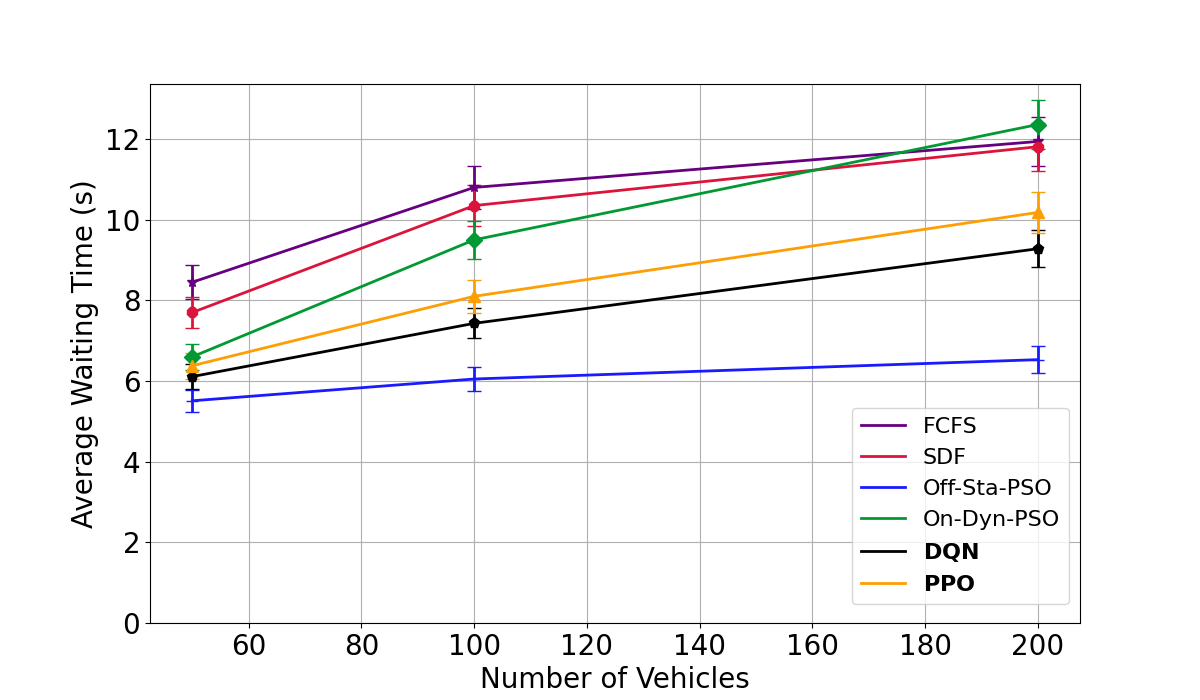}
    \caption{Average Waiting Time}
    \label{fig:waiting}
\end{figure}

\section{Performance Analysis}

\subsection{Experimental Setup}

In this study, a highway scenario is simulated using the SUMO platform, where a RSU facilitates computational offloading for vehicles traveling along the roadway. The RSU is equipped with two MEC servers to handle tasks generated by vehicles within its coverage area. Each task is associated with a time-sensitive deadline, calculated based on the Euclidean distance between the vehicle and the RSU. The simulation assesses the MEC servers' performance across three different traffic density scenarios to analyze the impact of varying workloads. Vehicles generate individual image-processing tasks, with task generation modeled using a Poisson distribution to reflect the variability typical in real-world vehicular offloading scenarios. Tasks can be initiated either while the vehicle is within the RSU's range or before entering it, enhancing the realism of the simulation. The size of each task is determined by the resolution of the processed image, with varying resolutions used to simulate different computational loads.

Data size calculations are based on a color depth of 24 bits per pixel, adhering to the standard RGB encoding format \cite{inference}. Simulation, PSO, and RL model parameters are configured based on \cite{mahsaicc, parisa2, dqnpara, ppopara1, ppopara2}, summarized in \tablename~\ref{tab:parameters}.

\subsection{Numerical Results}
\figurename~\ref{fig:reward} illustrates the reward values for DQN and PPO over 2500 training episodes.
Both DQN and PPO begin with rewards near 10,000, but PPO initially declines while DQN remains steady. After 1000 episodes, DQN progresses more rapidly, achieving higher rewards than PPO. Overall, DQN delivers stronger and more consistent performance. Moreover, \figurename~\ref{fig:allobjective} shows the optimization objective function values for Test-only both DQN and PPO, alongside other algorithms for comparison. In a dynamic setting, only the final sum of the best objective values from each run is reported. Similarly, since Test-only RL models are deployed for a single episode, their objective values remain constant.

DQN demonstrates stronger performance, achieving an objective value closest to that of Off-Sta-PSO (theoretical limit). All key performance metric values are averaged across 10 runs. The variation across runs indicates that even slight differences in decision-making can result in different tasks being dropped or processed, leading to significant changes in E2E latency and waiting time. \figurename~\ref{fig:drop} presents the drop task ratio for all algorithms under varying numbers of vehicles. The results show that Test-only DQN performs close to the theoretical limit and outperforms both On-Dyn-PSO and PPO.

Similarly, the average E2E latency results are shown in \figurename~\ref{fig:latency}. This indicates that Test-only DQN continues to outperform On-Dyn-PSO, which struggles to maintain performance under a high number of vehicles. This is primarily due to the high execution time of On-Dyn-PSO. Compared to PPO, DQN results in lower E2E latency as well. Due to the small magnitude of the original execution times, their logarithmic values are presented in \figurename~\ref{fig:execution} for more effective comparison. The values show that RL methods are faster than Dynamic PSO, which reduces the extra waiting time added due to algorithm execution, and as a result, leads to fewer dropped tasks. Among the RL approaches, DQN runs faster than PPO and also achieves better overall performance, making it the best trade-off between speed and effectiveness. 

To facilitate a better comparison between the models, the average waiting times are presented in \figurename~\ref{fig:waiting} as well. The performance trends are similar to those observed for E2E latency, as the waiting time constitutes the primary component of E2E latency. Models with longer waiting times inevitably exhibit higher E2E latency. As demonstrated, DQN achieves a lower average waiting time compared to PPO and dynamic PSO, positioning it closest to the theoretical limit. These results are also reported for First Come First Served (FCFS) and Shortest Deadline First (SDF), serving as benchmarks to enable a more comprehensive evaluation of the proposed models. Execution time details for Test-only PPO and DQN models can be found in \tablename~\ref{tab:exe}.

\section{Conclusion}
This study addresses the challenge of real-time task offloading in vehicular networks by evaluating both heuristic and RL models. The key contributions encompass the development of a theoretical performance benchmark using static PSO, the introduction of a dynamic decision window framework aimed at minimizing dropped tasks and reducing latency, and the implementation of DQN and PPO models with adaptive reward mechanisms to facilitate policy transfer without the need for retraining.
The proposed DQN model  significantly outperforms the On-Dyn-PSO, achieving a 99.2\% reduction in execution time, as well as 18.6\% lower E2E latency and 2.5\% fewer dropped tasks. Notably, the DQN model achieves performance that is closest to the optimal benchmark in terms of both E2E latency and the drop task ratio. Compared to PPO, DQN offers a 57.1\% decrease in execution time, as well as 5.7\% lower E2E latency and 1.7\% fewer dropped tasks. DQN outperforms PPO because it handles discrete, window-based task selection more efficiently, while PPO is better suited for continuous action spaces and requires more samples to stabilize. 

In our future work, we plan to consider deploying additional models that require no training and offer faster execution, and explore task partitioning techniques using RL methods. In addition, we will extend the current framework by incorporating Meta-RL to improve adaptability across varying task patterns and MEC configurations. While the current focus is on minimizing latency and reducing the number of dropped tasks, future efforts will broaden the optimization objectives to include energy consumption. We also aim to investigate the impact of scaling the number of MEC servers on the overall system performance, particularly its effect on the objective function involving delay and energy metrics.


 

\section*{Acknowledgment}
This work was supported in part by funding from the Innovation for Defence Excellence and Security (IDEaS) program from the Department of National Defence (DND), in part by Natural Sciences and Engineering Research Council of Canada (NSERC) CREATE TRAVERSAL Program, in part by Ontario Research Fund-Research Excellence (ORF-RE) program under RE012-026, and in part by the NSERC DISCOVERY Program.
\bibliographystyle{IEEEtran}

\end{document}